\documentclass[pre,nofootinbib,superscriptaddress,showpacs,showkeys]{revtex4}

\usepackage{graphicx,epsfig}
\usepackage{amsfonts,amsmath,amssymb,amsthm,amscd}
\usepackage[T2A]{fontenc}

\begin{document}

\title{Ensemble of ultra-high intensity attosecond pulses from laser-plasma interaction.}

\author{S. S. Bulanov}
\address{FOCUS Center and Center for Ultrafast Optical Science,
University of Michigan, Ann Arbor, Michigan 48109, USA}
\address{Institute of Theoretical and Experimental Physics,
Moscow 117218, Russia}

\author{A. Maksimchuk}
\address{FOCUS Center and Center for Ultrafast Optical Science,
University of Michigan, Ann Arbor, Michigan 48109, USA}

\author{K. Krushelnick}
\address{FOCUS Center and Center for Ultrafast Optical Science,
University of Michigan, Ann Arbor, Michigan 48109, USA}

\author{K. I. Popov}
\address{Theoretical Physics Institute, University of Alberta,
Edmonton T6G 2J1, Alberta, Canada}

\author{V. Yu. Bychenkov}
\address{P. N. Lebedev
Physics Institute, Russian Academy of Sciences, Moscow 119991,
Russia}
\address{Theoretical Physics Institute, University of Alberta,
Edmonton T6G 2J1, Alberta, Canada}

\author{W. Rozmus}
\address{Theoretical Physics Institute, University of Alberta,
Edmonton T6G 2J1, Alberta, Canada}

\begin{abstract}
The  efficient generation of intense X-rays and $\gamma$-radiation
is studied. The scheme is based on the relativistic mirror concept,
{\it i.e.}, a flying thin plasma slab interacts with a
counterpropagating laser pulse, reflecting part of it in the form of
an intense ultra-short electromagnetic pulse having an up-shifted
frequency. In the proposed scheme a series of relativistic mirrors
is generated in the interaction of the intense laser with a thin
foil target as the pulse tears off and accelerates thin electron
layers. A counterpropagating pulse is reflected by these flying
layers in the form of an ensemble of ultra-short pulses resulting in
a significant energy gain of the reflected radiation due to the
momentum transfer from flying layers.
\end{abstract}


\maketitle

\section{Introduction.}

{\noindent} The development of sources of intense ultra-short
electromagnetic (EM) pulses, X-rays, and even $\gamma$-rays is an
important potential application of intense laser matter interactions
\cite{atto}. Such applications vary from single molecule imaging to
radiography of dense targets and medicine diagnostics, and provide
the opportunity to study the fundamental effects not available
before, such as the effects of nonlinear Quantum Electrodynamics
(QED). One of the most promising ways to generate such sources is
the use of a reflection of EM radiation from a flying relativistic
mirror. This was first studied by Einstein in \cite{Einstein} as an
example of Lorentz transformations. The radiation frequency up-shift
is proportional to the square of the mirror Lorentz factor, making
the scheme very attractive for the generation of high frequency
pulses.

The principal idea of the relativistic plasma mirror has existed for
a long time \cite{BulanovNaumovaPegoraro}. Recently several ways to
create such mirrors have been proposed. One way is to use the plasma
waves in the wakefield of a high intensity pulse as it travels
through low density plasma in the wave breaking regime
\cite{EsirkepovTajimaBulanov}. The incoming light is reflected by
these waves in the form of a compressed pulse with an up-shifted
frequency (demonstrated in the experiment reported in Ref.
\cite{Pirozhkov2007}). Moreover additional intensification comes
from the parabolic form of the wake wave
\cite{Bulanov_Wake,MatlisNature}, which focuses the reflected
radiation. The role of the counterpropagating electromagnetic pulse
can be provided by different nonlinear structures in plasma, left in
the wake of another pulse, such as solitons, electron vortices, or a
wakefield. In this case the breaking plasma wave will reflect a part
of their electromagnetic energy in the form of a single-cycle
ultra-short high frequency pulse \cite{Isanin}. Another potential
method is the interaction of intense linearly polarized
electromagnetic pulses with solid density plasma, where either
sliding \cite{Pirozhkov2006} or oscillating mirrors
\cite{BulanovNaumovaPegoraro,OscillatingMirror1,OscillatingMirror2,OscillatingMirror3}
can be formed. The part of the incoming pulse that is reflected
carries high order harmonics due to the oscillation of the
reflecting surface either in the transverse or longitudinal
direction
\cite{Pirozhkov2006,OscillatingMirror1,OscillatingMirror2,OscillatingMirror3}.

It was proposed recently that relativistic mirrors can be formed in
the regimes of laser-thin foil interaction previously considered in
regard with the ion acceleration: in the first case the laser pulse
is intense enough to separate the electrons from the ion core, so
that the electrons will move in front of the pulse, forming a dense
relativistic electron layer. The counterpropagating laser pulse upon
reflection from such a mirror will undergo compression and frequency
up-shift \cite{Meyer-ter-Vehn,Habs}. In the second case a
double-sided relativistic mirror is formed. One side is used for the
energy transfer from the laser pulse that accelerates the foil.
While the other side can be utilized to reflect the incoming
radiation to produce ultra-short high frequency pulses
\cite{RPDmirror}.

In this letter we propose a more realistic mechanism of generation
of ultra-short EM pulses in laser-solid density target interaction,
then previously discussed in the literature
\cite{Meyer-ter-Vehn,Habs}. Sice it represents a more general
scenario of ultra-short pulse generation in laser-thin foil
interaction. In the proposed scheme the ultra-short EM pulses are
created in the course of a counterpropagating laser pulse
interaction with a series of flying electron layers, \textit{i.e.}
the Relativistic Multilayer Reflection (RMR) mechanism. Such layers
are produced when an intense laser pulse interacts with a thin solid
density target and extracts and accelerates thin electron layers
\cite{Naumova, Ma, Tian,PopovKI}. The high density and relativistic
velocity of these electron layers make it possible that such
structures will reflect the incoming radiation, acting as flying
mirrors. Since the incident pulse experiences the Relativistic
Multilayer Reflection, such interaction will result in the
generation of an ensemble of ultra-short pulses. Moreover this
scheme can lead to a significant energy gain by reflected radiation
due to the momentum transfer from flying layers. In what follows we
first present the results of 2D PIC simulation of the proposed
mechanism and then a 1D analytical model is employed to analyze the
results.

\section{The results of 2D PIC simulations.}

In our numerical model with the 2D PIC code REMP -- relativistic
electromagnetic particle - mesh code based on the particle-in-cell
method \cite{Esirkepov_code} the generation of ultra short pulses by
reflection from the series of laser accelerated dense electron
layers is studied in high-intensity laser interaction with
ultra-thin targets. The targets are composed of fully ionized carbon
C$^{+6}$ with an electron density of $400 n_{cr}$. The grid mesh
size is $\lambda$/200, space and time scales are given in units of
$\lambda$ and $2\pi/\omega$, respectively, the simulation box size
is $15\lambda\times 12.5\lambda$, where $\lambda$ and $\omega$ are
high-intensity laser wavelength and frequency respectively. The
number of particles per cell is 225. The $1.6$ PW laser pulse which
generates flying electron layers is introduced at the left boundary
and propagating along x axis from left to right. The pulse is
linearly polarized along the y axis (P-polarization), tightly
focused ($f/D=1$), and has Gaussian transverse and longitudinal
profiles. The duration of the pulse is 30 fs. The counterpropagating
laser, which is reflected from the relativistic mirrors, is
introduced at the right boundary and propagates from right to left
along the x axis. It is polarized along the z axis (S-polarization)
in order to distinguish between the radiation generated by the
accelerating pulse and the reflected one. It has $a_0=1$,
$\lambda_0=4\lambda$, and duration of 15 fs.

Below we present the results of 2D PIC simulations for the cases of
a mass limited target and a thin foil (see Fig. 1). In the first
case the counterpropagating laser pulse is reflected from the flying
electron layers accelerated from a disk with diameter $1\lambda$ and
thickness $0.1\lambda$ placed at $x=6.0\lambda$, \textit{i.e.}
before the focus of the accelerating pulse, which is focused at
$x=8.0\lambda$ (Figs. 1a-c). Such configuration is chosen in order
to show the formation of flying electron layers without a transverse
flow of electrons towards the irradiated spot and obtain a more
regular group of flying electron layers. In the second case the
pulse interacts with a $0.1\lambda$ thick foil placed at the focus
($x=6\lambda$) of the accelerating pulse. The electron density
distributions at $t=20$ are shown for both cases in Figs. 1a and 1d.
The thin electron layers that act as flying relativistic mirrors can
clearly be seen. The density of these layers vary from $5$ to
$15n_{cr}$ for the mass limited targets and from $10$ to $20n_{cr}$
for the thin foil. The duration of these bunches is about $70$ as.

\begin{figure}[ht]
\begin{tabular} {ccccc}
\epsfxsize5cm\epsffile{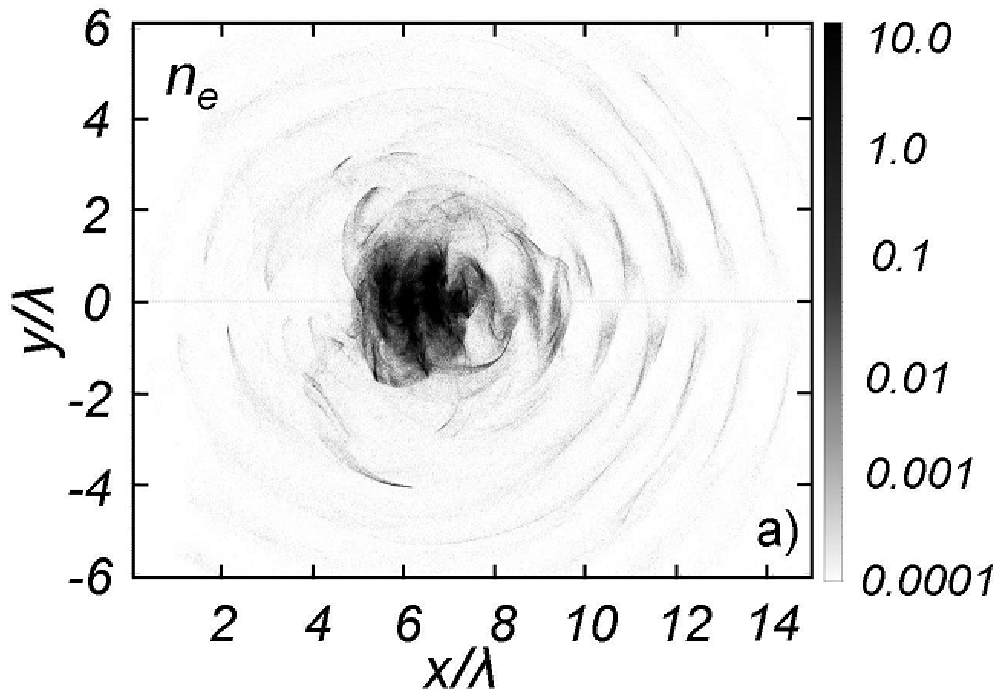}& & \epsfxsize5cm\epsffile{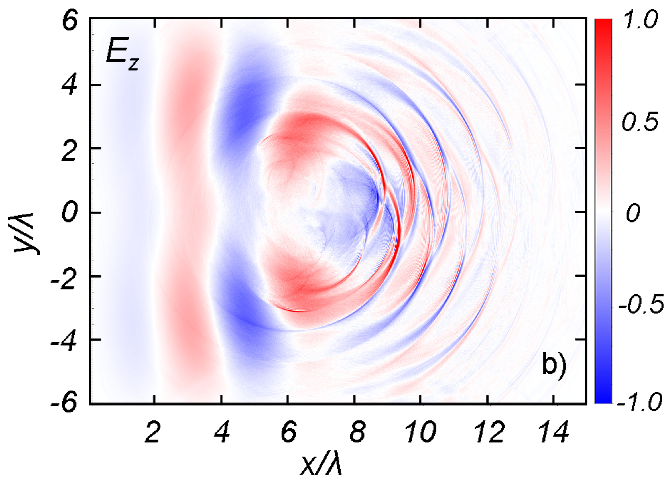}& & \epsfxsize5cm\epsffile{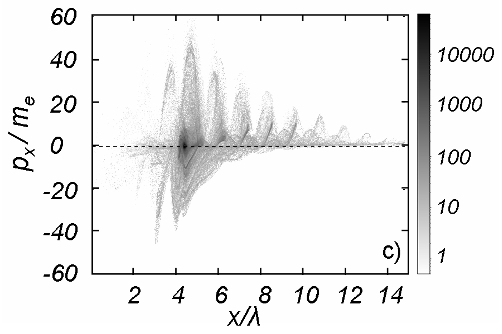}\\
\epsfxsize5.5cm\epsffile{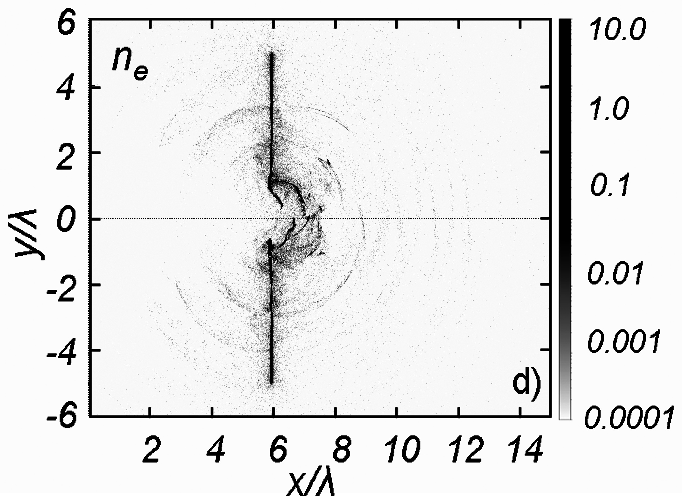}& &
\epsfxsize5.5cm\epsffile{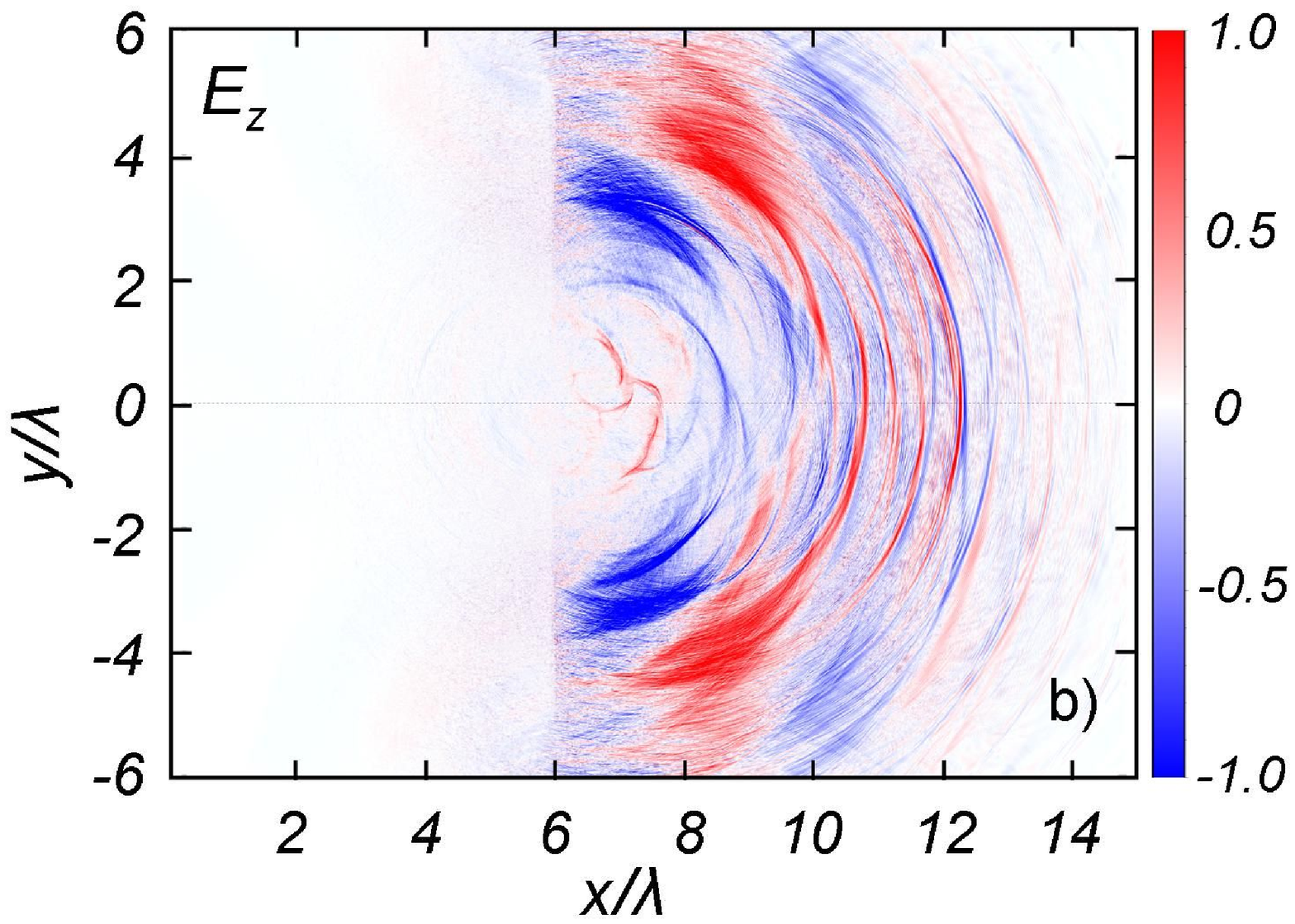}& & \epsfxsize5cm\epsffile{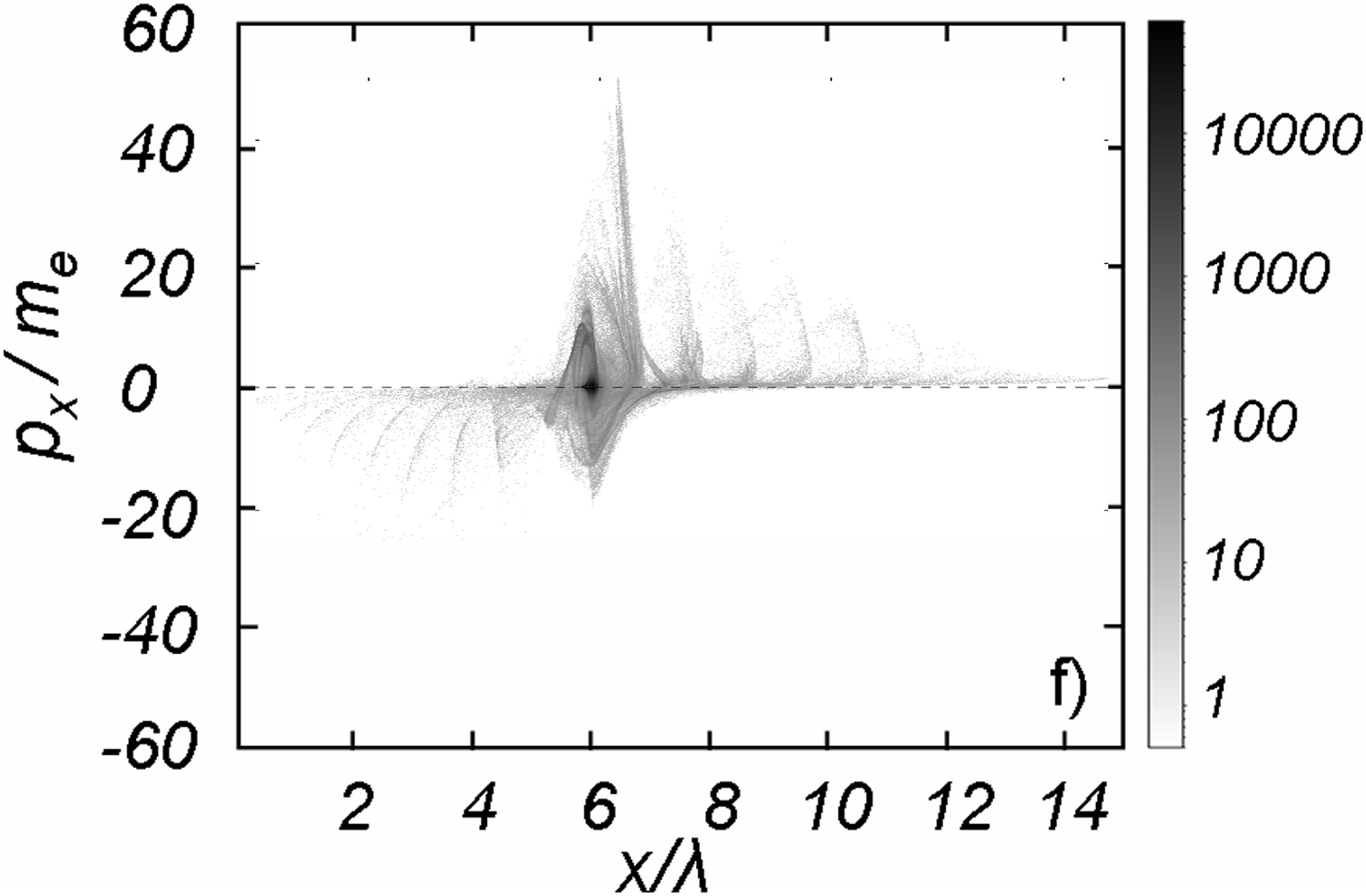}
\end{tabular}
\caption{(color on-line)The
reflection of laser pulse by accelerated electron layers in the case
of a mass limited target (upper row: a, b, and c) and a thin foil
(lower row: d, e, and f). The electron density distribution a) and
d) at $t=18$; the distribution of counterpropagating pulse electric
field after reflection, b) and e) at $t=18$, the field is measured
in units of $m_e c\omega/e$; the distribution of electrons in
($p_x,x$) phase plane at $t=12$, c) and f).}
\end{figure}

\section{1D model of relativistic mirror generation and radiation reflection by it.}

Let us estimate the properties of the flying electron layer using a
simple 1D model. As we have seen from the results of 2D PIC
simulations the electron layers are generated by the consecutive
maxima and minima of the laser electric field. The number of
electrons per flying layer can be obtained from the requirement that
the Coulomb attraction force should be greater than the Lorentz
force exerted by the EM field on these electrons. Since all the
extracted electrons almost instantly become relativistic, $v_e=c$,
the Lorentz force is proportional to laser electric field. Or in
terms of fields, the laser field should be greater than the charge
separation field, $E_{cs}=\pi e \delta n_e l$, which is generated
after the extraction of electrons ($l$ is the foil thickness,
$\delta n_e$ is the density of evacuated electrons). This gives the
condition $a>\pi \delta n_e l/n_{cr}\lambda$, first introduced in
Ref. \cite{OscillatingMirror2} as a condition for thin foil
transparency. Here $a$ is the amplitude of a vector potential in the
corresponding maxima or minima of the field, $n_{cr}$ is the
critical plasma density, and $\lambda$ in the laser wavelength. Here
we should note that the density of evacuated electrons is $\delta
n_e$ if $\delta n_e<n_e$, otherwise all the electrons are evacuated
and $\delta n_e=n_e$.

However, as it was pointed out in \cite{PopovKI}, this condition
does not take into account the fact that each electron layer,
escaping the attraction of the ion core, increases the charge
separation field that should be compensated for by the laser pulse
field. Let us approximate the field of the laser as $a_0
\exp[-t^2/\tau^2]\cos[2\pi t/T]$, where $\tau$ is the half of the
duration of the pulse and $T$ is the period of the EM wave. The
maxima and minima of such a wave are at $t_j=T j/2$, $j=0,\pm 1,\pm
2,...$. Then for some field maximum $a_j$ the charge separation
field that already exists is determined by $a_{j+1}$. Then the
number of electrons evacuated is $ \Delta N_j^e=\lambda R^2 n_{cr}
(a_j-a_{j+1})$, here $R$ is the radius of the irradiated area. The
total number of evacuated electrons will be $\Delta N^e=\sum_j
\Delta N^e_j=\lambda R^2 n_{cr} a_0$, \textit{i.e.} is determined by
the maximum of the vector potential only. Here we also assume that
$\Delta N_e<N_e$, otherwise $\Delta N_e=N_e$.

Let us estimate the duration of an electron bunch. The extraction of
electrons begins only when $a>a_{j+1}$ and stops at the top of the
current laser cycle $a=a_j$. Then the time interval $\xi$ that
determines the duration of the bunch can be found by solving the
equation $a(t_j+\xi_j)=a_{j+1}$:
\begin{equation}
\frac{\xi_j}{T}=\frac{1}{2\pi}\arccos\left\{\exp\left[-\frac{T^2}{\tau^2}
\left(\frac{j}{2}+\frac{1}{4}\right)\right]\right\}.
\end{equation}
The duration is minimal at the maximum of the pulse ($j=0$)
$\xi_0/T=0.11(T/\tau)$ and increases with the increase of $j$. For a
laser pulse with $\tau=5T$ and $T=3$ fs the electron bunch has an
attosecond duration: $\xi_0=60$ as, which is in an agreement with
the results of 2D PIC simulations as well as the number of electrons
per bunch.

Such thin flying electron layers can reflect the conterpropagating
radiation in the form of short pulses with up-shifted frequency. The
generation of ensembles of such EM pulses through the RMR mechanism
is demonstrated in Figs 1b and 1e. In order to determine the
properties of flying electron layers and determine Lorentz factors
of mirrors the distributions of electrons in ($p_x,x$) phase plane
are shown in Figs. 1c and 1f for $t=12$. The formation of flying
relativistic mirrors with $\gamma\sim 5$ can clearly be seen.

In order to characterize the generation of ultra-short EM pulses we
show the results of spatial Fourier analysis of reflected (curve 1)
and incident (curve 2) radiation in Fig. 2 for the case of the mass
limited target. The calculations are preformed for the field along
the line $y=0.6\lambda$ at $t=15$, when the maximum frequency
radiation can be resolved on our grid. As the interaction evolves we
expect further frequency up-shifting based on our theoretical model.
It can be clearly seen that high-frequency radiation is generated as
a result of counterpropagating pulse reflection from flying electron
layers. The spiky structure of the spectrum can be a consequence of
the regular distribution of reflecting electron layers in space (see
Fig. 1a).

\begin{figure}[ht]
\epsfxsize8cm\epsffile{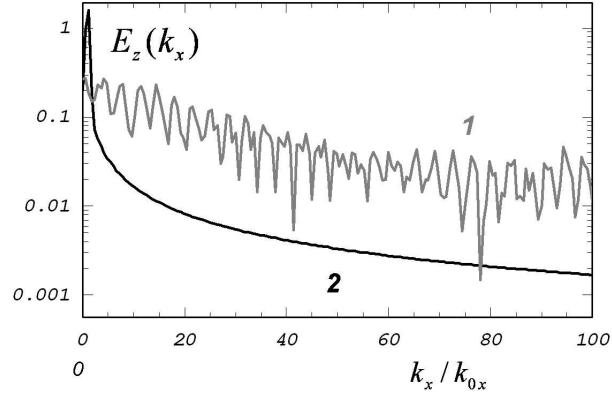} \caption{(color on-line)
The results of spatial Fourier analysis of the reflected (1,red) and
incident (2,black) pulses in the case of a mass limited target.}
\end{figure}

Let us estimate the reflection coefficient of the flying electron
layer to determine the efficiency of short pulse generation
mechanism. In order to do so we perform Lorentz transformation to
the reference frame co-moving with the electron layer and use the
results of Ref.\cite{OscillatingMirror2}, where the interaction of
an EM wave with a thin foil was considered and the reflection
($\rho=\epsilon_j/(i+\epsilon_j)$) and transmission
($\tau=i/(i+\epsilon_j)$) coefficients were obtained. Here
$\epsilon_j= \epsilon_0^j/(1+\beta)\gamma=\epsilon_0^j/2\gamma$
(since $\beta\sim 1$) is the parameter governing the transparency of
the foil in the co-moving with the foil frame and $\epsilon_0^j=\pi
\Delta n_j^e\xi_j/n_{cr}T$ is the transparency parameter of the
electron layer in the laboratory frame \cite{OscillatingMirror2}.
Since it is essential for our mechanism that the flying electron
layers remain unperturbed by counterpropagating pulse we take the
intensity of such pulse to be relatively small.

Let us estimate the intensification of the laser pulse reflected by
such a mirror. For the incident laser intensity, $I_0$,  the
reflected pulse in the laboratory frame will have the up-shifted
frequency by a factor of $4\gamma^2$ and the increased intensity
\begin{equation}
I=\frac{16\epsilon_0^{j~2}\gamma^4}{4\gamma^2+\epsilon_0^{j~2}}I_0
=\left\{
\begin{tabular}{c}
$16\gamma^4 I_0,~~~\gamma\ll \epsilon_0^j$\\ \\
$4\epsilon_0^{j~2}\gamma^2I_0,~~~\gamma\gg\epsilon_0^j$
\end{tabular}
\right. .
\end{equation}
or in terms of energy
\begin{equation}
\mathcal{E}_r=\frac{4\epsilon_0^{j~2}\gamma^2}{4\gamma^2+\epsilon_0^{j~2}}\mathcal{E}_0
=\left\{
\begin{tabular}{c}
$4\gamma^2 \mathcal{E}_0,~~~\gamma\ll \epsilon_0^j$\\ \\
$4\epsilon_0^{j~2}\mathcal{E}_0,~~~\gamma\gg\epsilon_0^j$
\end{tabular}
\right. .
\end{equation}
For $\gamma\ll \epsilon_0^j$ the reflection coefficient tends to
unity and reflected intensity end energy are determined by the
Lorentz factor alone. In the second case, $\gamma\gg\epsilon_0^j$,
the foil moves so fast that it becomes increasingly transparent for
incoming radiation and the energy of the reflected pulse is limited
by the transparency parameter. Thus the efficiency of the light
intensification is determined by
$\gamma^2\times\min\{\gamma^2,\epsilon_0^{j~2}\}$ for intensity and
$\min\{\gamma^2,\epsilon_0^{j~2}\}$ for energy. This means that the
maximum intensity increase is determined by the properties of the
accelerated foil. In order to illustrate the results of the
theoretical model we chose $\epsilon_0^j\sim 10$ and $\gamma\sim 10$
as parameters to estimate the reflected pulse intensity. These
numbers come from the results of 2D PIC simulations. The frequency
up-shift in this case is $4\times 10^2$ and for $T=3$ fs the period
of the reflected radiation will be $T_r=7.5$ as. Then the intensity
is $I\sim 3\times 10^{22}$ W/cm$^2$.

As we have shown the energy of the reflected from the flying
electron layer EM pulse is limited by $\epsilon_0^{j~2}$, if we
consider the dependence on $\gamma$. However there is a way to
increase the energy of reflected radiation by utilizing the fact
that the accelerating laser pulse extracts multiple electron layers
from the target. Moreover this type of reflection is realized in the
results of 2D PIC simulations presented above (see Figs 1a and 1d).
The counterpropagating pulse experience relativistic multilayer
reflection giving rise to an ensemble of ultra-short pulses. While
the number of photons reflected at each layer, $\Delta
N_\gamma^r=|\rho|^2 N_\gamma^0$ ($N_\gamma^0$ is the number of
photons in the counterpropagating pulse) can be small, the multiple
layers will reflect almost all incoming photons,
$N_\gamma^r\rightarrow N_\gamma^0$. If the number of layers is $f$,
then the total number of reflected photons \cite{Panchenko} is
\begin{equation}
N_\gamma^r=N_\gamma^0|\rho|^2\sum\limits_{j=1}^f
(1-|\rho|^2)^j=N_\gamma^0\left[1-(1-|\rho|^2)^f\right].
\end{equation}
Here we assumed for simplicity that all electron layers are the
same. Then in terms of energy
\begin{equation} \label{energy gain}
\mathcal{E}_r=4\gamma^2\left[1-(1-|\rho|^2)^f\right]\mathcal{E}_0.
\end{equation}
If $|\rho|^2 f\ll 1$ then $\mathcal{E}_r=4\gamma^2 |\rho|^2
f\mathcal{E}_0$. In the limit $f\rightarrow\infty$ the reflected
back energy equals to $\mathcal{E}_r=4\gamma^2\mathcal{E}_0$ and
$N_\gamma^r=N_\gamma^0$, \textit{i.e.} all the counterparpagating
radiation is reflected. The energy gain is due to the momentum
transfer from the flying electron layers to the reflected radiation.
For the parameters of the electron layers obtained in simulations
(the average density and duration of about $10 n_{cr}$ and 60 as,
the number of layers is equal to 4) the reflection coefficient is
0.05 and the energy gain is 1.5, according to (\ref{energy gain}).
We can estimate the actual energy gain in the results of 2D PIC
simulations from the spectra of incident and reflected radiation by
integrating $|E_z(k_x)|^2$ over $dk_x$. This gives a 1.07 energy
gain. This clearly shows that the proposed mechanism of RMR
generation of ultra-short pulses leads to an energy gain of
reflected radiation due to the momentum transfer from flying
mirrors.

\section{Towards the Schwinger field.}

Further intensification can possibly be achieved by focusing by some
external means the reflected pulse into a diffraction limited spot.
For a single pulse it will lead to
\begin{equation}
I_f\simeq\left(\frac{D}{\lambda_r}\right)^2 I= \left\{
\begin{tabular}{c} $\displaystyle{256\gamma^8 \left(\frac{D}{\lambda_0}\right)^2
I_0,~~~\gamma\ll\epsilon_0^{j}}$ \\ \\ $\displaystyle{64\gamma^6
\epsilon_0^{j~2} \left(\frac{D}{\lambda_0}\right)^2 I_0,~~~\gamma\gg
\epsilon_0^j}$
\end{tabular}
\right.
\end{equation}
where $D$ is the reflected pulse width before focusing and
$\lambda_r=\lambda_0/4\gamma^2$. Then for the parameters of the
flying layer used above, $D=3\lambda_0$ and $I_0\sim 10^{18}$
W/cm$^2$ the resulting intensity will be of the order of the
intensity characteristic for the effects of nonlinear QED,
\textit{i.e.} Schwinger intensity, $I_S\sim 10^{29}$ W/cm$^2$,
\cite{Schwinger}. At this intensity the probability of one of the
most profound processes of nonlinear QED, the $e^+e^-$ pair
production in vacuum by strong EM field, becomes optimal. We should
note here that the plane EM wave does not produce pairs in vacuum
\cite{Schwinger}, because in this case both field invariants,
$\mathcal{F}=\mathbf{E}^2-\mathbf{H}^2$,
$\mathcal{G}=\mathbf{E}\mathbf{H}$, are equal to zero, which is not
the case for the focused pulse \cite{NBPM}. Thus the focusing of
high frequency pulses to ultra-high intensity will provide a unique
opportunity to study both the $e^+e^-$ pair production by a focused
EM pulse and its dependence on frequency of the pulse \cite{Popov}.

\section{Conclusions.}

In this letter we considered a new way to generate ultra bright high
intensity X-rays and gamma-rays by reflecting EM pulse from the
relativistic mirror. In the proposed scheme the role of the flying
mirror is taken by laser accelerated electron layers, which are
formed in the process of the intense laser pulse interaction with
thin solid density targets. The reflected pulses have an up-shifted
frequency and increased intensity. The reflected pulse
intensification is determined by a combination of two parameters:
the Lorentz factor of the flying electron layer, $\gamma$, and its
transparency parameter, $\epsilon_0^{j}$, \cite{OscillatingMirror2}.
It is proportional to
$\gamma^2\times\min\{\gamma^2,(\epsilon_0^{j})^2\}$ and for given
density of the plasma slab the energy of the reflected pulse
multiplication factor can not exceed $(\epsilon_0^{j})^2$. It is due
to the fact that for $\gamma\gg \epsilon_0^j$ the fast moving plasma
slab becomes increasingly transparent for the incoming radiation and
the amplitude of the reflected pulse drops, limiting the energy
gain. Further intensification of the reflected light can possibly be
achieved by its focusing into a diffraction limited spot that will
bring the resulting peak intensity well into the domain of nonlinear
QED with laser systems, which are presently available.

We showed that there is a way to increase the energy transfer from
the accelerated layers to the reflected radiation by utilizing the
relativistic multilayer reflection. The counterpropagating EM pulse
interacts with multiple flying electron layers, producing an
ensemble of ultra-short pulses with an energy scaling of
$4\gamma^2$. This fact leads to the conclusion that this mechanism
of ultra-short pulse generation provides a highly efficient way of
transforming the incoming pulse into high frequency radiation
through the relativistic multilayer reflection.

This work was supported by the National Science Foundation through
the Frontiers in Optical and Coherent Ultrafast Science Center at
the University of Michigan and Russian Foundation for Basic
Research.

\end{document}